\documentclass[11pt,preprint2]{aastex} 
\newcommand\msun{\ensuremath{M_\sun}}
\newcommand\teff{\ensuremath{T_{\rm eff}}}
\newcommand\logg{\ensuremath{\log g}}
\newcommand\ubv{\ensuremath{U\!BV}}
\newcommand\ebv{\ensuremath{E(\bv)}}

\slugcomment{Accepted for publication in the Astrophysical Journal Letters} 
\shorttitle{A DQ white dwarf in NGC 2168}
\shortauthors{Williams et al.}

\begin{document}

\title{A Hot DQ White Dwarf in the Open Star Cluster M35} 

\author{Kurtis A.~Williams\altaffilmark{1}, James Liebert\altaffilmark{1}, 
Michael Bolte\altaffilmark{2}, and Robert B.~Hanson\altaffilmark{2}}

\altaffiltext{1}{Steward Observatory, 933 N. Cherry Ave., Tucson, AZ 85721}
\altaffiltext{2}{UCO/Lick Observatory, University of California, Santa
  Cruz, CA 95064} 

\begin{abstract}
We report the discovery of a hot DQ white dwarf, \object{NGC
2168:LAWDS 28}, that is a likely member of the 150-Myr old cluster
\object{NGC 2168} (Messier 35).  The spectrum of the white dwarf is
dominated by \ion{C}{2} features. The effective temperature is
difficult to estimate but likely $\gtrsim 20\,000\,{\rm K}$ based on
the temperatures of hot DQs with similar spectra.  NGC2168:LAWDS 28
provides further evidence that hot DQs may be the ``missing''
high-mass helium-atmosphere white dwarfs.  Based on published studies,
we find that the DBA WD LP 475-242 is likely a member of the Hyades
open cluster, as often assumed.  These two white dwarfs are the entire
sample of known He-atmosphere white dwarfs in open clusters with
turnoff masses $\geq 2\msun$.  Based on the number of known cluster DA
white dwarfs and a redetermination of the
H-atmosphere:He-atmosphere ratio, commonly known as the DA:DB ratio,
we re-examine the hypothesis that the H- to He-atmosphere ratio in
open clusters is the same as the ratio in the field.  Under this
hypothesis, we calculate that five He-atmosphere WDs are expected to
have been discovered, with a probability of finding fewer than three
He-atmosphere white dwarfs of 0.08, or at the $\approx 2\sigma$ level.
\end{abstract}

\section{Introduction}
White dwarfs (WDs) have one of two types of atmospheres:
those dominated by hydrogen (spectral class DA), and those dominated
by helium (the non-DA spectral classes).  The ratio of the numbers of
DA to non-DA stars is a function of temperature, culminating in the
``DB gap'', ranging from $\approx 45\,000\,{\rm K}$ to
$\approx 30\,000\,{\rm K}$ \citep[e.g.,][]{Liebert1986}. Until
recently, no He-atmosphere WDs were known to exist in this temperature
range, though several have now been discovered in the Sloan Digital
Sky Survey \citep{Eisenstein2006}.  The temperature dependence of the
DA to non-DA ratio suggests that some He-atmosphere WDs must
change their atmospheric composition, an effect often ascribed to the
diffusion of hydrogen to the surfaces of cooling DO stars and the
later mixing of this hydrogen veneer into the thicker He-envelope due
to the appearance of an atmospheric convection zone
\citep{Fontaine1987}.

\citet{Kalirai2005} note that even though a few dozen WDs are known to
exist in open star clusters with turnoff masses greater than 2\msun,
no DB WDs are known in these clusters.  However, this study excludes
other WD spectral types with He-dominated atmospheres, such as the
DBA WD LP~475-242 (WD 0437+138), a potential member of the
Hyades. \citet{Kalirai2005} also give only a brief mention of the
impact of the DB gap without calculating the impact of the changing DA
to non-DA ratio in this temperature range. Subsequent to their
publication, analysis of the SDSS DR4 WD sample \citep{Eisenstein2006}
has found that the DA to non-DA ratio is larger than the value of 3.5
used in \citet{Kalirai2005}, warranting a re-analysis of the cluster
non-DA deficiency.

In this Letter, we report the discovery of a hot DQ (He-dominated
atmosphere with opacity dominated by atomic carbon) WD in the field of
the young open cluster \object{NGC 2168} (Messier 35).  Several DA WDs
in this cluster have already been identified and characterized
\citep{Williams2004}. The identification of this He-atmosphere WD
provides insight into both the origin of DQ WDs and the apparent
deficiency of He-atmosphere WDs in open clusters.

\section{Observations\label{sec.obs}}
\begin{figure}
\plotone{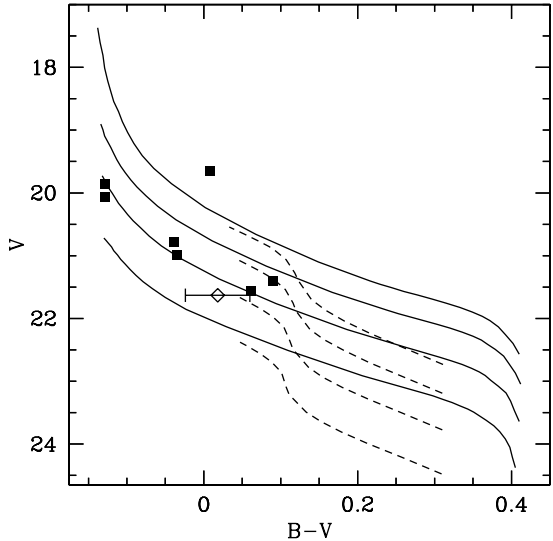}
\caption{Color-magnitude diagram of confirmed white dwarfs in NGC
  2168.  Curves are evolutionary tracks for pure H- (i.e., DA; solid)
  and pure He-atmosphere (i.e., DB; dashed) white dwarfs hotter than
  $10\,000$ K with masses of 0.6\msun, 0.8\msun, 1.0\msun, and
  1.2\msun (top to bottom, respectively).  Tracks have been moved to
  the cluster distance and reddening.  Filled squares are the
  confirmed cluster member white dwarfs from \citet{Williams2004}.
  The open diamond with $1\sigma$ error bars is NGC 2168:LAWDS 28; the
  error in $V$ is smaller than the point size.  This white dwarf has
  photometry consistent with known cluster white dwarfs and with
  massive white dwarf evolutionary models.
\label{fig.cmd}}
\end{figure}

We obtained \ubv~imaging of the open cluster NGC 2168 on UT 2004
January 27 with the Mosaic-1 camera on the Kitt Peak Mayall 4-m
telescope.  Photometry was derived via point-spread function fitting
using the DAOPHOT II program \citep{Stetson1987}.  We select objects
with blue excess as candidate WDs as discussed in
\citet{Williams2004}.  The color-magnitude diagram for confirmed
cluster WDs is shown in Figure \ref{fig.cmd}. NGC 2168 has an age of
$150\pm 60$ Myr \citep{vonHippel2005} and $[{\rm Fe}/{\rm H}]
=-0.21\pm 0.10$ \citep{Barrado2001}.  The main-sequence turnoff mass
is $\sim 4.5\msun$ \citep{Girardi2002}; based on the initial-final
mass relation, WDs in NGC 2168 are expected to be massive
\citep[e.g.,][]{Weidemann2000,Ferrario2005}.

The open diamond in Figure \ref{fig.cmd} is \object{NGC 2168:LAWDS 28}
(hereafter LAWDS 28), $\alpha(J2000)=6^{\rm h}08^{\rm m}13\fs 50$,
$\delta(J2000)=+24\degr 20\arcmin 32\farcs 5$, and we assign it the
\citet{McCook1999}-style designation WD J0608+2420.  This object has a
$V$-magnitude of $21.63\pm 0.03$, $\bv=0.02\pm 0.04$, and $\ub =
-1.10\pm 0.04$.  As its photometric properties (open diamond in Figure
\ref{fig.cmd}) are similar to those of known WDs in M35, we
selected this object for follow-up spectroscopy.

The spectroscopic data were obtained on UT 2005 November 26 using the
blue camera of the Low-Resolution Imaging Spectrometry on the Keck 1
telescope \citep{Oke1995}.  We placed a 1\arcsec~slit on the object at
the parallactic angle and selected the 400 line mm$^{-1}$,
3400\AA~blaze grism to obtain a spectral resolution of $\approx 6$\AA.
Our total exposure time was 4800s, split into four 1200s
exposures. The images were combined and the spectrum was extracted and
calibrated (wavelength and relative flux) using standard \emph{IRAF}
routines.

The extracted spectrum is peculiar and markedly different from
other spectra obtained in our open cluster WD survey, both for
confirmed WDs (DA, DB and DCs) and the non-WD sample contaminants
(mostly QSOs, hot subdwarfs, and field horizontal branch
stars).  The spectral features are consistent with \ion{C}{2} spectral
features, leading us to conclude that this object is a DQ WD.

\begin{figure}
\plotone{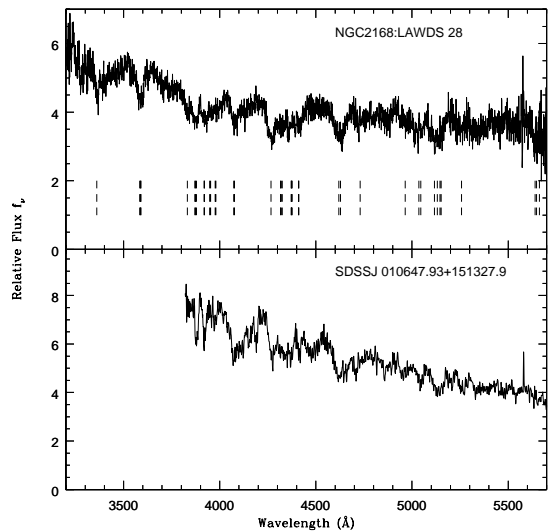}
\caption{Spectrum of NGC 2168:LAWDS 28 (top) compared to the known hot
DQ SDSS J010647.9+151327.9 (bottom). Dashed tick marks indicate the
location of \ion{C}{2} features.  The qualitative match of the two
spectra and the fact that most or all of the features in the top
spectrum match the \ion{C}{2} features lead us to conclude that
NGC2168:LAWDS 28 is a hot DQ white dwarf.\label{fig.spec}}
\end{figure}

The spectrum for LAWDS~28 is shown in Figure \ref{fig.spec} (top),
along with the SDSS 2.5-m spectrum for
\object{SDSS~J010647.93+151327.9} (bottom), the object most similar
from the sample of hot DQ stars showing \ion{C}{1}, \ion{C}{2},
\ion{O}{1} and/or \ion{O}{2} multiplets in \citet{Liebert2003}.  The
tick marks label the approximate positions of multiplets of \ion{C}{2}
taken from \citet{Moore1970}.  Included are the weaker multiplets
omitted from the plots in \citet{Liebert2003}.  Both multiplets with
strong and weak laboratory intensities appear in LAWDS 28; the 4267\AA
~line, often observed to be strong in hot stars, is present in
LAWDS 28 but is not the only dominant feature.  While a case could be
made that the SDSS object shows additional features, possibly due to
\ion{O}{2} in its spectrum, the cluster WD is matched almost perfectly
with \ion{C}{2} transitions alone.  It is possible that the strong
\ion{C}{3} multiplet near 4647--51\AA ~may contribute weakly, but no
\ion{C}{1} or oxygen lines are matched except where they would blend
with the \ion{C}{2}.

Hot DQ WDs are thought to have atmospheres dominated by helium.
\citet{Thejll1990} analyze the atmospheric abundances of the hot DQ
\object{G35-26} (WD 0203+207) and find a carbon abundance $n_{\rm
C}\sim 0.03$ and a hydrogen abundance of $n_{\rm H}\lesssim 0.01$,
implying that $n_{\rm He}\sim 0.96$ \citep{Liebert2003}.  The hot DQ
\object{G227-5} (WD 1727+560) also was found to have only trace
abundances of carbon and hydrogen \citep{Wegner1985}.  Therefore it is
likely that LAWDS 28 has a He-dominated atmosphere despite showing no
evidence for helium features in the spectrum.  The lack of such
features in many hot DQs is thought to be due to the masking of helium
features by the dominant carbon opacity \citep{Liebert2003}.

\section{Discussion\label{sec.disc}}

\subsection{Cluster membership, mass, and temperature}
Given the lack of a model fit to the spectrum of LAWDS 28, it is
difficult to determine many fundamental parameters of this WD with
precision.  However, the available observations provide a
preponderance of evidence that LAWDS 28 is a hot, massive
cluster-member WD.  We now outline this evidence.

The distance and reddening of NGC 2168 are somewhat uncertain.
\citet{Kalirai2003} adopt $\ebv = 0.20$ from \citet{Sarrazine2000} to
derive $(m-M)_0 = 9.80\pm 0.16$ from $BV$ photometry, while
\citet{Sung1999} calculate $(m-M)_0=9.60\pm 0.1$ and $\ebv = 0.255 \pm
0.024$ from $\ubvr I$ photometry.  Either case leads to an apparent
distance modulus $(m-M)_V\approx 10.4$.  At this distance, the
absolute magnitude of LAWDS 28 is $M_V=11.2$; assuming the
\citet{Sarrazine2000} reddening value, $(\bv)_0 = -0.22$ and $(\ub)_0
= -1.23$.  This absolute magnitude is consistent with evolutionary
models of DA and DB WDs with $M_{\rm WD}\gtrsim 0.8\msun$ and hotter
than $20\,000 {\rm K}$ \citep{Bergeron1995}.  However, DQ photometric
indices are likely to differ from those of DAs and DBs, and in fact
the hot DQs in \citet{Liebert2003} are bluer in $u-g$ than pure-He
atmosphere models.  As seen in Figure \ref{fig.cmd}, LAWDS 28 is also
bluer than the given DB models.

The effective temperature, \teff, of LAWDS 28 cannot be determined 
accurately, as no spectral models for these hot DQs exist.  The \bv
~color of this WD is bluer than any DB model from
\citet{Bergeron1995}, though these models are not calculated for
$\teff>30\,000\,{\rm K}$.  By matching the observed \bv ~color to
reddened blackbody models calculated using the \emph{synphot} package
in STSDAS \citep{Bushouse1994}, we derive a blackbody temperature of
$22\,200^{+5200}_{-3500}\,{\rm K}$.  For the hot DQs in
\citet{Liebert2003}, SDSS colors were converted to a ``pure helium''
effective temperature, $T_{\rm He}$, based on colors for pure
helium-atmosphere WDs with $\logg=8$.  For SDSS J0106+1513, the
comparison star shown in Figure \ref{fig.spec}, $T_{\rm He} =
20\,000\,{\rm K}$, while the DQs \object{SDSS J0005-1002} and
\object{SDSS J0236-0734} have $T_{\rm He} = 29\,000$ and
$30\,000\,{\rm K}$, respectively \citep{Liebert2003}.  Both of these
stars show \ion{C}{1} in addition to \ion{C}{2} in their spectra; the
lack of \ion{C}{1} in the spectrum of LAWDS 28 would argue that this
DQ is hotter than the other two. While the continuum slope of SDSS
J0106+1513 is bluer than that of LAWDS 28, this does not necessarily
mean that LAWDS 28 is cooler, as the spectra have not been corrected
for extinction.  Since neither the blackbody temperature nor the
pure-helium temperatures take the spectral features of the DQs into
account, these are only educated guesses as to \teff.  Given the very
blue colors and the ion ratios, we surmise that LAWDS 28 is a hot
($\teff\gtrsim 20\,000\, {\rm K}$) WD.  We also note that the coolest
DA WDs in NGC 2168 (similar in \bv ~to LAWDS 28) have $\teff \sim 30\,000\,{\rm K}$.

The discovery of a hot DQ white dwarf in this young cluster is
consistent with the case reiterated in \citet{Liebert2003} that such
WDs may be massive.  \citet{Thejll1990} found a high gravity for the
hot DQ star \object{G 35-26} from their detailed atmospheric analysis
and fit to the spectral lines of carbon, hydrogen and helium.  This
has been confirmed by a trigonometric parallax measurement
\citep{Dahn2006}.  A parallax is also available from the same source
for the similar field star \object{G 227-5}, the spectrum of which was
analyzed by \citet{Wegner1985}, and it also appears to have a mass
near 1~\msun.  It is expected from theoretical evolutionary
calculations of asymptotic giant branch stars that the helium-shell
burning consumes a greater fraction of the helium layer with
increasing mass \citep{Kawai1988}, with further calculations leading
to a similar conclusion in \citet{Thejll1990}.  A smaller helium
envelope as the star enters the white dwarf cooling phase in turn
means that dredge-up of carbon (and perhaps oxygen) from the edge of
the core into the convective envelope and atmosphere should occur at a
higher temperature -- in the DB temperature range, in which LAWDS~28
apparently is.  If \emph{all} hot DQ stars are in fact massive, this
may account for the absence of massive DB stars in the sample of
\citet[][see also \citealt{Beauchamp1996}]{Beauchamp1995}; this is the
only large sample of field DB stars for which detailed determinations
of the relevant parameters (\teff, \logg, and therefore mass) are
available. Nearly all of these DB stars have masses near the
canonical peak of 0.6 \msun.

Finally, hot DQs are very rare.  In the SDSS DR4 WD catalog
\citep{Eisenstein2006b}, only 0.3\% (28 out of 9316 white dwarfs) are
hot DQs; with a surface density of approximately one for every 180
square degrees.  We estimate the number of hot white dwarfs in our 0.5
sq.~degree imaging of NGC 2168 by integrating the WD luminosity
function from the Palomar-Green Survey \citep{Liebert2005} for
$M_V\geq 11$ to a volume corresponding to limiting apparent magnitude
of $V=21.6$.  This estimate therefore allows for the possibility that
LAWDS 28 is less massive than 1\msun, and therefore brighter and
background to NGC 2168.  The estimated number of intrinsically bright
WDs in our imaging area is found to be $\approx 10$.  If 0.3\% of
these are hot DQs, we would expect 0.03 hot DQs in our imaging area.
Thus, the likelihood that LAWDS 28 is an interloping field DQ in our
0.5 square degree field of view is extremely low.  However, until a
distance modulus or proper motion can be determined for LAWDS 28, some
uncertainty about its membership will remain.

In summary, the photometric and spectroscopic data for LAWDS 28 are
consistent with it being a hot ($\teff \gtrsim 20\,000\,{\rm K}$),
massive ($M\gtrsim 0.8\msun$) WD located at the distance of NGC 2168.
Until an accurate model atmosphere can be fit to the data or until
proper motion information is available, it is not possible to
state \emph{conclusively} that LAWDS 28 is a member of NGC 2168.

\subsection{Other non-DA WDs in open clusters}

\emph{LP~475-242. ---} LP 475-242 (WD 0437+138) is a DBA white dwarf
about one core radius (3 pc) from the Hyades cluster center.
\citet{Beauchamp1995} fits its spectrum to get $\teff =
15\,355\,{\rm K}$ and $\logg = 8.26$, suggesting a modestly
above-average mass ($0.751\pm 0.033\msun$) for a DB white dwarf.  
Hydrogen lines were included to get a trace hydrogen abundance of
$\log (n_{\rm H}/n_{\rm He}) = -4.5$, so the atmosphere is clearly
He-dominated.  This WD has been used in some cluster data compilations
to determine the initial-final mass relation, most recently that of
\citet{Ferrario2005}.

LP 475-242 was listed by \citet{Luyten1971} as a possible white dwarf
member of the Hyades on the basis of its proper motion, faint
magnitude, and blue color.  \citet{Greenstein1976} determined a
photometric parallax of 0\farcs 023, in close agreement with the
Hyades distance.  \citet{Eggen1985} and \citet{Weidemann1992} consider
LP 475-242 to be an established Hyades member.  Accurate proper motion
measurements (Table \ref{tab.pm}) agree almost exactly, in size
and direction, with motion expected for a Hyades member.  Finally,
under the assumption that LP 475-242 is at the systemic velocity of
the Hyades, the resulting gravitational redshift leads to a mass of
$0.748\pm 0.037\msun$ \citep{Reid1996}, consistent with the
spectroscopic mass \citep{Beauchamp1995} and with the masses of DAs in
the Hyades \citep{Bergeron1995a}.  It therefore seems highly likely
that LP 475-242 is a Hyades member.

\emph{LB 3600. ---} It is also possible that a third cluster DB is
already known, \object{LB 3600} in \object{Messier 67}
\citep{Fleming1997}.  Based on their spectral fits, this WD is about
one magnitude too bright to be a member of M67, though their spectrum
is very noisy, and DB spectra are difficult to fit with precision.  As
the deficiency of open cluster DBs noted by \citet{Kalirai2005} was
specifically for clusters with main-sequence turnoff masses $\geq
2\msun$ and the M67 turnoff mass is $\approx 1.4\msun$, we do
not consider this object in our analysis below.

\subsection{The open cluster DA:non-DA ratio}

\citet{Kalirai2005} note the lack of DB WDs in open clusters.
Though there are no known ``pure'' DB WDs in open clusters, this
overlooks the physical reality that non-DA spectral types in the DB
temperature range -- DBs, DBAs, DBZs, and hot DQs -- have atmospheres
dominated by He.  As He has a very low opacity, trace contaminants in
the atmosphere can easily dominate the opacity, masking He 
features.  In order to study whether the ratio of hydrogen-dominated
atmosphere WDs to helium-dominated atmosphere WDs differs from star
clusters to the field, these other spectral types should be included
in any analysis.  Given our discovery of LAWDS 28, a likely
He-dominated atmosphere WD in an open cluster, given the probable
Hyades DBA LP~475-242, and given a new determination of the field
DA:DB ratio from Sloan Digital Sky Survey data \citep{Eisenstein2006},
we now revisit the issue of the H- to He-atmosphere ratio in open
clusters.

\emph{A new measurement of the field DA:DB ratio. ---} The large
numbers of field WDs discovered during the Sloan Digital Sky Survey
has permitted a new determination of the DA:DB ratio (including DBAs)
\citep{Eisenstein2006}.  For $\teff\lesssim 20\,000\,{\rm K}$,
DA:DB$\approx 5$, while for $20\,000\,{\rm K} \lesssim \teff\lesssim
45\,000\,{\rm K}$, DA:DB$\approx 12.5$.  Because of the near
degeneracy of DB spectra over the temperature range of
$20-28\,000\,{\rm K}$, it is difficult to determine the
lower \teff~boundary of the deficit, though it is most
likely in the $20-22\,000\,{\rm K}$ range.  Thus, the
``DB-gap'' is really a deficiency, not a complete lack, of WDs in the
temperature range $45\,000\,{\rm K} \lesssim \teff \lesssim
20\,000\,{\rm K}$, somewhat larger than previous estimates of the DB
gap range.  As this is a large, uniform sample of WDs, and as the
DA:DB ratio is provided as a function of temperature, we adopt these
ratios for calculating the number of expected He-atmosphere WDs in the
open cluster sample.

\emph{Expected numbers of He-atmosphere WDs in the open cluster
sample. ---} For the sample of open cluster WDs, we use the sample as
presented by \citet{Ferrario2005}.  This work applies stringent
membership criteria, so few non-cluster members should contaminate the
sample.  We calculate the number of non-DA WDs using the
\citet{Eisenstein2006} DA:DB ratio in two temperature bins: $\teff <
20\,000\,{\rm K}$ (DA:DB$=5$; 21 open cluster DAs) and $20\,000\leq
\teff \leq 45\,000\,{\rm K}$ (DA:DB$=12.5$; 17 DAs).  We thus expect
$\approx 5.6$ non-DA WDs in the current open cluster WD sample.  This
number assumes that the \citet{Eisenstein2006} DA:DB ratio would be
unaffected by inclusion of non-DB and non-DBA He-atmosphere WDs in the
calculations.

Given two probable He-atmosphere cluster WDs (LAWDS 28 and
LP~475-242), we use Poisson statistics to calculate the likelihood of
finding two or fewer non-DA WDs to be 0.08.  If only one of these
He-atmosphere WDs is a cluster member, this probability drops to 0.02.
This is akin to $1.9\sigma$ and $2.3\sigma$, respectively.  Thus,
while the ratio of DA to non-DA WDs in open clusters is low, it is not
greatly inconsistent with the expected number.  With coming larger
samples of open cluster WDs, the question of whether the DA to non-DA
ratio in open clusters is the same as that in the field can be
addressed with much more certainty.

\acknowledgements The authors are grateful for financial support from
National Science Foundation grants AST 03-07492 and AST 03-07321.  The
authors also thank the referee, Harvey Richer, for comments improving
this paper.

\begin{deluxetable}{cccccl}
\rotate
\tablewidth{0pt}
\tablecolumns{6}
\tablecaption{Proper motion measurements of LP 475-242.\label{tab.pm}}
\tablehead{\colhead{$\mu_\alpha$} & \colhead{$\mu_\delta$} &
  \colhead{$\mu$} & \colhead{PA} & \colhead{Reference} & \colhead{Comments}}
\startdata
0.093\tablenotemark{a} & -0.036\tablenotemark{a} & 0.10 & 111  & (1) & \\
0.100 & -0.021 & 0.102\tablenotemark{a} & 102\tablenotemark{a} & (2) & \\
0.097 & -0.023 & 0.100\tablenotemark{a} & 103\tablenotemark{a} & (3) & LP 475-242 = Reid 400 \\
0.094 & -0.018 & 0.096\tablenotemark{a} & 101\tablenotemark{a} & (4) & LP 475-242 = NOMAD 1039-0046673 \\
0.098\tablenotemark{a} & -0.021\tablenotemark{a} & 0.100\tablenotemark{a} & 102\tablenotemark{a} & (5) & Predicted motion for a Hyades member at position of LP 475-242 \\
\enddata
\tablenotetext{a}{Calculated from data in reference}
\tablecomments{Proper motions are in arcseconds per year; position angles
are degrees east of north.}
\tablerefs{(1) \citealt{Luyten1971}, (2) \citealt{Luyten1981}, 
  (3) \citealt{Reid1992},  (4) \citealt{Zacharias2004}, (5) \citealt{Perryman1998}}
\end{deluxetable}

\end{document}